# Electronic structure of confined carbyne from joint wavelength-dependent resonant Raman spectroscopy and density functional theory investigations

*Miles Martinati[a], Wim Wenseleers[a*], Lei Shi[b], Saied Md Pratik[c], Philip Rohringer[a,d], Weili Cui[d], Thomas Pichler[d], Veaceslav Coropceanu[c], Jean-Luc Brédas[c], Sofie Cambré[a*].*

[a] Nanostructured and Organic Optical and Electronic Materials group, Physics Department, University of Antwerp, Universiteitsplein 1, B-2610 Antwerp, Belgium

[b] State Key Laboratory of Optoelectronic Materials and Technologies, Nanotechnology Research Center, Guangzhou Key Laboratory of Flexible Electronic Materials and Wearable Devices, School of Materials Science and Engineering, Sun Yat-sen University, Guangzhou 510275, P. R. China

[c] Department of Chemistry and Biochemistry, The University of Arizona, Tucson, Arizona 85721-0088, United States

[d] Faculty of Physics, University of Vienna, Strudlhofgasse 4, A-1090 Vienna, Austria

* corresponding authors: wim.wenseleers@uantwerp.be and sofie.cambre@uantwerpen.be.

**ABSTRACT**

Carbyne, *i.e.* an infinitely long linear carbon chain (LCC), has been at the focus of a lot of research for quite a while, yet its optical, electronic, and vibrational properties have only recently started to become accessible experimentally thanks to its synthesis inside carbon nanotubes (CNTs). While the role of the host CNT in determining the optical gap of the LCCs has been studied previously, little is known about the excited states of such ultralong LCCs. In this work, we employ the selectivity of wavelength-dependent resonant Raman spectroscopy to investigate the excited states of ultralong LCCs encapsulated inside double-walled CNTs. In addition to the optical gap, the Raman resonance profile shows three additional resonances. Corroborated with DFT calculations on LCCs with up to 100 C-atoms, we assign these resonances to a vibronic series of a different electronic state. Indeed, the calculations predict the existence of two optically allowed electronic states separated by an energy of 0.14-0.22 eV in the limit of an infinite chain, in agreement with the experimental results. Furthermore, among these two states, the one with highest energy is also characterized by the largest electron-vibration couplings, which explains the corresponding vibronic series of overtones.

**KEYWORDS**: carbyne, Raman spectroscopy, resonance Raman profiles, excited states, vibrational overtones, density functional theory.

1. **Introduction**

An infinitely long linear carbon chain (LCC) or carbyne, consisting of a single string of sp-hybridized carbon atoms, represents the truly 1D allotrope of carbon. In addition to excellent mechanical properties in terms of stiffness, strength, and elastic modulus[1], LCCs show a remarkably high Raman cross-section, in particular when encapsulated inside carbon nanotubes (CNTs)[2]. It is well recognized that the electronic and optical properties of the LCCs directly correlate with their bond-length alternation (BLA)[3–6] pattern. The presence of BLA follows from Peierls' theorem: Since an infinitely long 1D chain of equally spaced carbon atoms would correspond to a metallic electronic structure, Peierls' theorem establishes that this situation is unstable *vis-à-vis* a distortion of the lattice leading to an alternation of single and triple bonds (polyynic structure) and the opening of a bandgap at the Fermi level.[3] The BLA is associated with a Raman active phonon mode, called the C-mode, that corresponds to the in-phase stretching of the triple bonds (and associated shrinking of the single bonds; in other words a modulation of the BLA) along the chain. The vibrational frequency of this C-mode sensitively depends on the extent of BLA, *i.e.*, the higher the BLA, the higher its frequency.[7,8] Since BLA in relatively short LCCs is a critical function of chain length and nature of the chemical groups at their ends,[6,9] as well as extrinsic effects such as strain, doping[4,10] and environment[11], the bandgap and vibrational properties of the LCCs can be tuned. This feature is appealing for a broad variety of potential applications, such as highly sensitive temperature sensors.[12] On the other hand, the extreme dependence of the LCC properties on a broad range of parameters, makes their prediction very difficult.

While the synthesis of long LCCs has long remained elusive since the high reactivity of the chains limited lengths up to 48 carbon atoms[13,14], increasingly longer LCCs have been realized via their synthesis inside CNT cavities.[15–18] These cavities act both as a confined nano-reactor for the synthesis, thereby preventing larger structures to be formed, and as a protecting barrier to the environment, overcoming the instability problem of the chains in free space. Lengths up to 6000 contiguous atoms have been reported upon synthesis inside the thin inner cavity of double-walled CNTs (DWCNTs).[18] With such long lengths available, the properties of the chains have been found to no longer depend on length, indicating that they are the finite realization of carbyne, the infinitely long LCC.[6,19]

Raman spectroscopy (RS) is particularly useful for the characterization of LCCs encapsulated inside CNTs, due to the extremely high Raman cross-section of the LCCs, which are found to be the strongest Raman scatterers ever reported.[2] Moreover, resonant Raman spectroscopy (RRS) provides simultaneous access to both the vibrational and electronic properties of the encapsulated chains as well as to the properties of the host CNTs. As mentioned above, the Raman spectrum of LCCs shows only one first-order Raman-active mode, the C-mode.[6,9,17] Previously, it was demonstrated that the frequency of this C-mode for ultralong LCCs inside DWCNTs depends on the diameter of the CNT host, while no longer a function of chain length, proving that the BLA and its properties have reached the limit of an infinitely long LCC, hence called carbyne.[19] In fact, the smaller the CNT diameter, the lower the Raman frequency of the C-mode.[19] In a macroscopic sample comprising many different CNT chiralities and diameters, this results in a C-mode composed of a discrete number of Raman frequencies, in the range of 1790-1860 $cm^{-1}$.[6,18] By monitoring the Raman intensity of these modes as a function of laser excitation wavelengths, it was found that also the optical gap of these encapsulated LCCs depends on the CNT diameter, as a result of the interaction with the host CNT due to the confinement.[6] Similarly, wavelength-dependent Raman spectroscopy has been used to measure the optical gap of individual LCCs.[20]

While in short LCCs in solution the absorption spectra of the chains consist of multiple resonances that are assigned to different electronic and vibrational excited states[14,21,22], the vibronic fine structure for ultralong CNT-encapsulated chains has remained elusive to date with only the optical gap having been reported in detail. One study by Fantini *et al.*[23] did present the resonance Raman profile of short encapsulated LCCs; however, the authors were unable to resolve and assign the different components of the C-mode to a specific Raman profile and thus only investigated the overall resonance Raman profile (RRP) of the combined C-modes.

In this work, we combine a high spectral resolution and a broad tunable laser excitation range to measure wavelength-dependent RRS for LCCs encapsulated in DWCNTs with different chiralities. This approach allows us to resolve the vibrational, vibronic, and electronic fine structure of ultralong LCCs confined inside DWCNTs with different inner tube chiralities and diameters, in a macroscopic ensemble sample. Our results are further corroborated by theoretical calculations indicating that the RRPs consist of multiple resonances that can be assigned to a combination of two optically allowed electronic transitions and a vibronic series of one of these optically allowed transitions.

## 2. Experimental details

### 2.1. Synthesis of the LCC@DWCNT samples

Two different CNT samples, prepared in the form of buckypapers, were utilized as hosts for the formation of confined LCCs. The first sample started from single-walled CNTs (SWCNTs), with an average diameter of ~1.3 nm, synthesized by the enhanced direct injection pyrolytic synthesis method (hereafter denoted as eDIPS-SWCNTs)[24]. The second sample directly started from DWCNTs, with an average outer diameter of around 1.5 nm, grown by chemical vapor deposition (hereafter denoted as CVD-DWCNTs).[18,25]  In order to synthesize the confined LCCs, a procedure similar to that in Ref. [[18]] was used. First, the eDIPS-SWCNTs and CVD-DWCNTs were annealed at 1460°C for 1 hour under high vacuum (below $10^{-7}$ mbar). In this manner, the eDIPS-SWCNTs are transformed into DWCNTs by the formation of ultrathin inner tubes after annealing.[26] Simultaneously to the formation of these inner SWCNTs, the confined LCCs are grown inside those new inner tubes[27]. This LCC@eDIPS sample will be denoted as Sample 1. For the as-grown CVD-DWCNTs, the inner tubes are already present and the confined LCCs are directly formed within these inner tubes. This LCC@CVD sample will be denoted as Sample 2. The different diameter distributions present within these two samples allow for a broader set of LCC@CNT configurations to be probed.

### 2.2. Wavelength-dependent resonant Raman spectroscopy

Wavelength-dependent RRS was performed directly on the bucky papers in the excitation wavelength range from 400 to 800 nm (3.10–1.55 eV) with 5-nm step size. To this end, a combination of tunable laser systems were used: (i) 400-526 nm from a single frequency Ti:Saphire laser with external-cavity frequency doubler (SolsTiS ECD-X platform from M Squared Lasers Limited), which was pumped with an 18W Sprout-G diode-pumped solid-state laser (532 nm) from Light-house Photonics; (ii) a dye laser (Spectra Physics model 375) pumped by an Ar$^+$ ion laser (Spectra Physics model 2020), with two laser dyes, either Rhodamine 110 (534-605 nm) or DCM (610-690 nm); and (iii) a tunable Ti:Saphire laser (Spectra Physics model 3900S) that was pumped by the same Ar$^+$ ion laser to cover the wavelength range from 690-800 nm.

For the Raman experiments, a high-resolution triple-grating Dilor XY800 Raman spectrometer was used, equipped with a liquid nitrogen cooled CCD detector and 1800gr/mm gratings. Raman intensities of the LCCs@DWCNTs samples were normalized by measuring the Raman intensity of the 520.7 cm$^{-1}$ Raman mode of silicon before and after each experiment (see more details on the intensity calibration further in the man text and Supporting Information section SI.2). The position of this peak was also used to correct for minor changes of the calibration of the spectrometer at each laser wavelength.

### 2.3. Details of the simultaneous Raman fits

The resonant Raman spectra acquired at different laser excitation energies are fitted simultaneously. To this end, we devised a fitting model that is composed of a sum of Lorentzians of which the peak positions and line widths are shared for all the Raman spectra within the RRS map of each sample. While the peak positions and line widths of these Lorentzians are obtained through a numerical least-squares fitting algorithm, their amplitudes at each laser excitation energy are calculated analytically by linear regression. As such, the fit is composed of a linear combination of basis functions corresponding to each C-mode (and the background), of which the line width and peak position are numerically optimized, and the coefficients of this linear combination are determined from the linear regression. This approach results first of all in a much better determination of the Raman frequencies and line widths of the different C-modes, since they are now defined by taking into account all the Raman spectra in which they are observed and since their relative amplitudes vary in between spectra due to the different resonance behavior of each of the modes. In addition, such a simultaneous fitting procedure allows us to determine with the highest precision the amplitudes of those modes in spectra with very low intensities. The spectra of the two samples are fitted separately, each time using a sum of 6 Lorentzian functions (thus 12 fit parameters) to account for the LCC Raman modes while a third-order polynomial function was previously subtracted to account for the background in this region. The error bars on the fit parameters (positions and line widths) were calculated considering the correlation between the parameters; for the amplitudes, the small variation in laser intensity during the experiments was also taken into account. Moreover, to account for imperfections of the fit model, which lead to non-normally distributed residuals around zero, we effectively reduced the number of degrees of freedom in the fit to the number of zero-crossings in the residuals, which is typically much lower. This effectively results then in larger error bars on the fit parameters, which thus accounts for systematic deviations between the model and the experimental data.

## 2.4. Computational details

The geometry optimizations and vibrational characterizations of hydrogen-terminated carbon chains ($C_nH_2$, where n=20, 30, 40, 50, 60, 70, and 100) were carried out at the density functional theory (DFT) level, using the long-range corrected ωB97X-D functional and the def2-TZVP basis set. Subsequent time-dependent DFT (TD-DFT) calculations were performed to compute the excited-state properties. In these calculations, the screened range-separated hybrid (SRSH) functional LC-ωhPBE and the 6-31+G(d,p) basis set were used. The range-separation parameter (ω) was optimally tuned for each chain through the minimization of the expression $J(\omega) = (E_{HOMO} + IP)^2 + (E_{LUMO} + EA)^2$, as described in detail elsewhere.[28–31] Here, $E_{HOMO}$ and $E_{LUMO}$ denote the HOMO and LUMO energies, while IP and EA denote the vertical first ionization potential and electron affinity of the chain. The SRSH-based calculations allow us to account effectively for the screening effect on the chains due to the solid-state environment.[32] Here, this was done by following common practice[32,33] and using dielectric constant ($\varepsilon$) values typical for organic systems in the range of 2-4. This dielectric screening effectively shifts the electronic transitions to lower energies, to an absolute value closer to the experimentally obtained values. Also, in order to estimate the electron-vibrational couplings, we computed the Huang-Rhys (HR) factors for the n=30 chain. All the DFT calculations were performed with the Gaussian 16 package.[34]

## 3. Results and Discussion

The complete RRS maps obtained for the two LCC@DWCNT samples are shown in Figure 1, where Raman intensity is plotted as a function of Raman frequency and laser excitation energy. Starting from low Raman frequencies, both samples show the radial breathing modes (RBMs) of the DWCNTs in the range between 100 and 400 cm$^{-1}$ with the typical splitting of the inner CNT RBMs due to their encapsulation in different outer CNTs.[35] From Figure 1, it can be observed that the diameter distributions differ for both samples, with Sample 1 presenting higher Raman intensities for smaller diameter inner tubes (*i.e.*, those with higher RBM frequencies). The dispersive D-mode of the CNTs can be found around 1300 cm$^{-1}$ and blue-shifts with increasing laser excitation energy, while the CNT G-band is centered around 1600 cm$^{-1}$. Finally, the characteristic Raman mode of the LCCs (*i.e.*, the C-mode) can be found in between 1790 and 1860 cm$^{-1}$ and shows several components for the two samples, which are resonant at different excitation energies. The difference in Raman frequency of the C-modes in the two samples can be related to the variations in diameter distribution between the two samples. Indeed, Heeg *et al.*[19] showed that the C-mode frequency decreases with decreasing SWCNT diameter, which can be explained by the higher confinement of the LCCs for such small-diameter SWCNTs; this is consistent with our observation that in Sample 1, lower-frequency C-modes (1790-1800 cm$^{-1}$) can indeed be observed with a higher relative amplitude than in Sample 2.

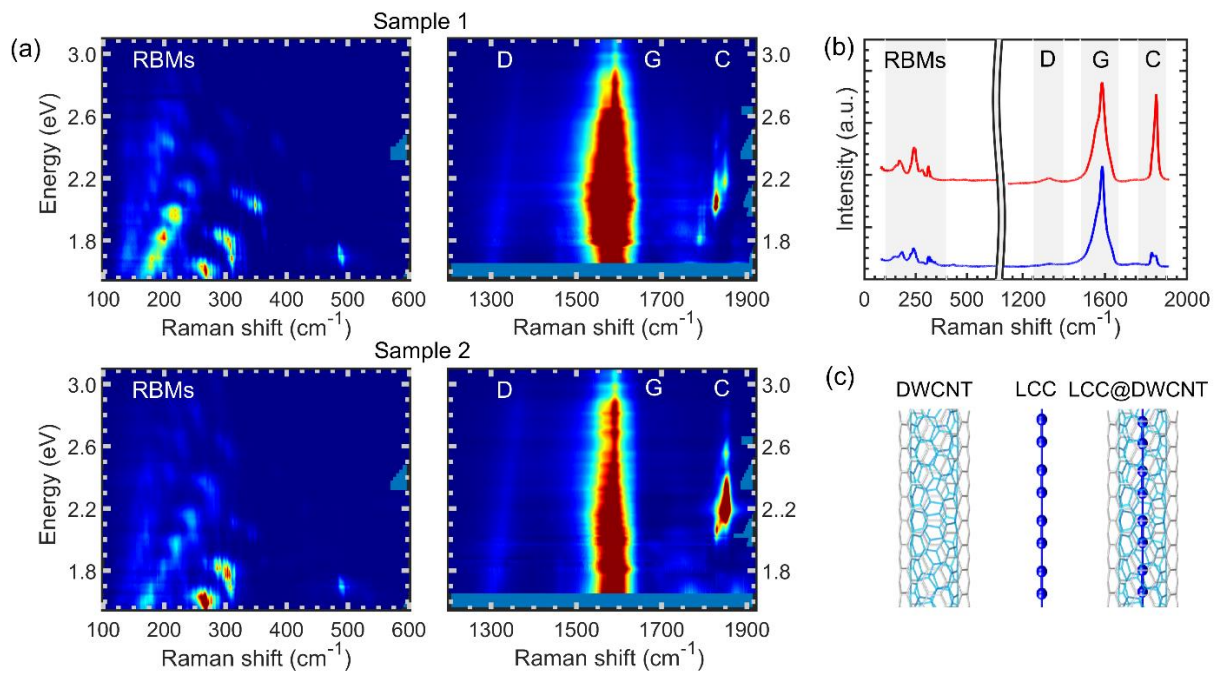

Figure 1: (a) Wavelength-dependent RRS of the two samples (top = Sample 1, bottom = Sample 2) of LCCs@DWCNTs. Raman intensity (colorscale) is plotted as a function of Raman shift and laser excitation energy in the ranges of the radial breathing modes (RBMs) of the DWCNTs (left panels) and the ranges of the D- and G-bands of the DWCNTs and the C-mode of the LCCs (right panels). From the RBM panels, it can be observed that Sample 1 contains more of the thinner inner SWCNTs (higher RBM frequencies) as compared to Sample 2, resulting also in the other components (at lower frequency) seen in the C-mode region. (b) Raman spectra at a laser excitation energy of 2.1 eV of Sample 1 (blue) and Sample 2 (red), vertically shifted for clarity. The Raman spectra at low Raman frequencies (RBM region) are multiplied by a factor of 4 for clarity. (c) Representation of a DWCNT, a LCC and of a LCC encapsulated inside a DWCNT.

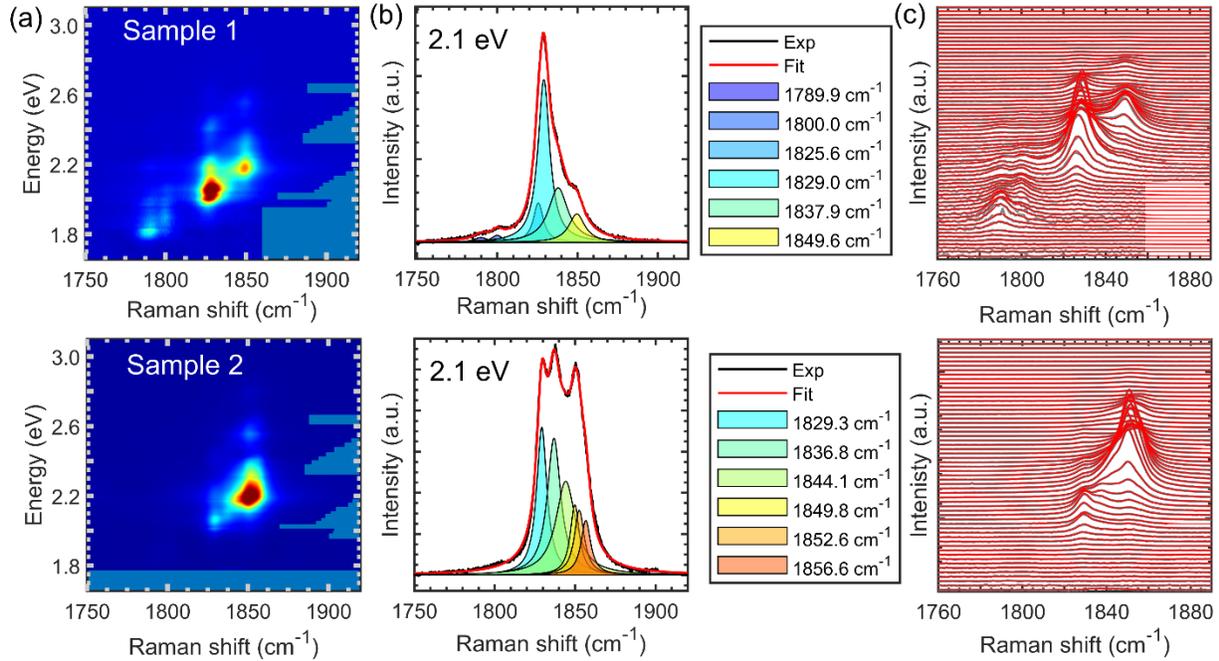

Figure 2: (a) Wavelength-dependent RRS maps of the LCCs@DWCNTs samples presented in Figure 1, zoomed into the C-mode region. The top panel represents Sample 1, the bottom panel Sample 2. (b) Individual Raman spectra of the two samples at a laser excitation energy of 2.1 eV. The experimental data are shown in black, the best fit is shown in red, while each component used in the simultaneous fit procedure is shown in shaded areas. (c) Raman spectra (grey dots) of the two samples, vertically shifted for different values of laser excitation energy, with the best fits (in red; almost perfectly coinciding with the experimental data), as obtained by the simultaneous fit.

The Raman spectra of the two samples at a laser excitation energy of 2.1 eV are presented in Figure 2b and evidence that the C-band is composed of many components, corresponding to LCCs encapsulated into different inner tubes of the host CNTs. Since all these Raman modes are very close in frequency and have line widths of approximately 10 cm$^{-1}$, resolving and assigning the different Raman modes is only possible by simultaneously fitting all the spectra at different laser energies within one RRS map, *i.e.*, with shared peak positions and line widths for all excitation energies, but allowing the amplitudes of the fitted peaks to vary for each excitation energy, as described in Section 2.3. A complete 2D fit of the Raman maps is presented in Figure 2c, highlighting the accuracy of this fitting procedure.

The spectra of the two samples are each fitted with two different sets of six Raman frequencies for the C-mode with three of them equal in both samples (1829, 1837 and 1850 cm$^{-1}$), for a total of nine different components overall ranging from 1790 to 1857 cm$^{-1}$. Sample 1 has in general higher intensities for lower C-mode frequencies compared to Sample 2. This agrees with the observed differences in diameter distribution in the RBM range. Indeed, the highest C-mode intensity in Sample 1 is observed for the C-mode with frequency 1830 cm$^{-1}$ which according to Ref. [19] corresponds to an LCC encapsulated inside the (6,5) chirality. The RBM of the (6,5) chirality is indeed much more present in Sample 1 as compared to Sample 2 (see Figure 1, RBM ≈ 310-320 cm$^{-1}$, laser energy ≈ 2.1 eV). Likewise, when extrapolating the observed diameter-dependence of the C-mode from Ref. [19], the highest LCC frequencies observed in Sample 2 should correspond to a surrounding (7,5)/(7,6) inner-tube chirality, of which the RBMs are indeed clearly present in Sample 2 (see Figure 1, RBM ≈ 280-310 cm$^{-1}$ , laser energy ≈ 1.8 eV). Raman frequencies and line widths of the C-mode components are listed in

Table 1 and are in good agreement with previous studies on the same kind of samples.[6,19] All the LCC Raman modes have line widths in between 7 and 15 cm$^{-1}$, which is (at least) an order of magnitude higher than the resolution of the Raman spectrometer used to measure these RRS maps, indicating that they represent the intrinsic line width of those C-mode components.

|  | Sample 1 | | | | | |
| --- | --- | --- | --- | --- | --- | --- |
| ν (cm$^{-1}$) | 1789.9 ± 0.6 | 1800.0 ± 0.4 | 1825.6 ± 0.1 | 1829.0 ± 0.1 | 1837.9 ± 0.4 | 1849.6 ± 0.1 |
| Δν (cm$^{-1}$) | 10.8 ± 0.1 | 8.8 ± 0.1 | 7.6 ± 0.1 | 8.9 ± 0.1 | 14.7 ± 1.0 | 12.1 ± 0.2 |
| E$_1$ (eV) | 1.825 ± 0.006 | 1.918 ± 0.021 | 2.008 ± 0.006 | 2.058 ± 0.003 | 2.129 ± 0.003 | 2.197 ± 0.005 |
| Γ$_1$ (meV) | 109 ±21 | 190 ± 47 | 100 ± 20 | 86 ± 9 | 131 ± 9 | 138 ± 16 |
|  | Sample 2 | | | | | |
| ν (cm$^{-1}$) | 1829.3 ± 0.1 | 1836.8 ± 0.1 | 1844.1 ± 0.1 | 1849.8 ± 0 | 1852.6 ± 0.3 | 1856.6 ± 0.1 |
| Δν (cm$^{-1}$) | 7.8 ± 0.1 | 10.1 ± 0.4 | 14.8 ± 0.3 | 6.9 ± 0.1 | 7.6 ± 0.2 | 7.4 ± 0.1 |
| E$_1$ (eV) | 2.061 ± 0.004 | 2.092 ± 0.003 | 2.179 ± 0.004 | 2.196 ± 0.004 | 2.217 ± 0.007 | 2.250 ± 0.006 |
| Γ$_1$ (meV) | 101 ± 11 | 77 ± 9 | 110 ± 11 | 106 ± 14 | 131 ± 22 | 146 ± 17 |

Table 1 Raman frequencies and vibrational line widths of the various components of the LCC C-mode for the two samples, combined with the energies and electronic line widths of the first resonance (optical gap) observed in the RRPs.

After fitting the 2D RRS maps, the amplitudes of each of the Lorentzian functions can be plotted as a function of the laser excitation energy, yielding the RRPs for each component of the C-band observed in the two samples. These RRPs are presented in Figure 3. In agreement with observations for shorter polyynes[14,21,22], each of the C-mode RRPs presents a series of four resonances with decreasing intensity at higher laser excitation energy. The lowest-energy and most intense Raman resonance corresponds to the optical gap, which was previously investigated by some of us in Ref. [6]. Three other resonances are also observed, whose origin will be discussed below. To extract their respective energies, the RRPs are fitted by a semi-classical resonant Raman model [36]:

$$I(E_L) = \sum_j \left| \frac{M_j}{E_L - E_j - i\Gamma_j} \right|^2 \qquad (1)$$

where $M_j$ is the incident resonance factor, $E_L$ the laser excitation energy, $E_j$ the energy of each resonance, $\Gamma_j$ the corresponding broadening term, and the sum over $j$ takes into account the different resonances in each RRP. This fit function does not include the term in resonance with the emitted photon energy (the so-called out-going resonance). Note that the typical scattering factor into the intensity (proportional to the energy of the scattered light $E_S^4 = (E_L - E_{vib})^4$ with $E_L$ the laser energy and $E_{vib}$ the vibrational energy) is automatically eliminated by the calibration of the Raman intensity with the Si peak (see Section 2.3). However, since the energy of the scattered light is different in the LCCs compared to the Si sample, we have to additionally correct by a factor $(E_S^{Si}/E_S^{LCC})^4$. Finally, the Silicon Raman intensity is also dependent on the laser excitation energy [37], which should be taken into account to have a correct data normalization [38,39]. The RRPs are then corrected also for the dependence of the Raman intensity of Silicon (more information on this calibration procedure is reported in the SI). To obtain the energy, line width and amplitude of the fourth (highest energy) transition, it was fitted separately in a narrower energy range and afterwards these fit values were used as fixed parameters when fitting each RRP. The best fits of the RRPs are reported in red in Figure 3. The corresponding energies of these Raman resonances for each of the C-mode frequencies are presented in Table 1, as well as shown in Figure 4.

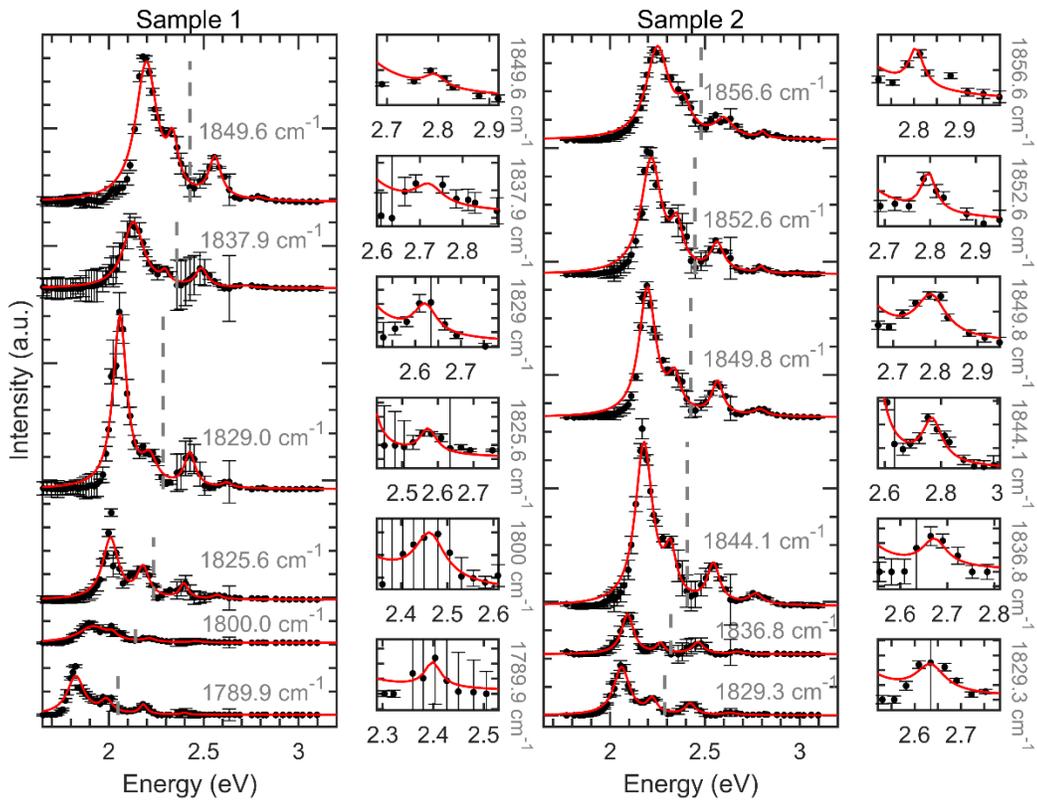

Figure 3: Amplitudes of the Lorentzian components (black circles), each with their respective error bars, and fitted resonance Raman profiles (RRPs, in red). While the amplitudes are obtained from fitting the C-mode RRS maps presented in Figures 1-2 with different Lorentzian components, the RRPs are fitted using equation (1) in the main text. The vertical grey dashed lines indicate the vibrational energy (from the Raman frequency of the C-mode) with respect to the first Raman resonance (R1). The RRPs in general show four resonances (R1-R4); the smaller insets zoom into the 4$^{th}$ highest energy resonance for each of the C-mode frequencies.

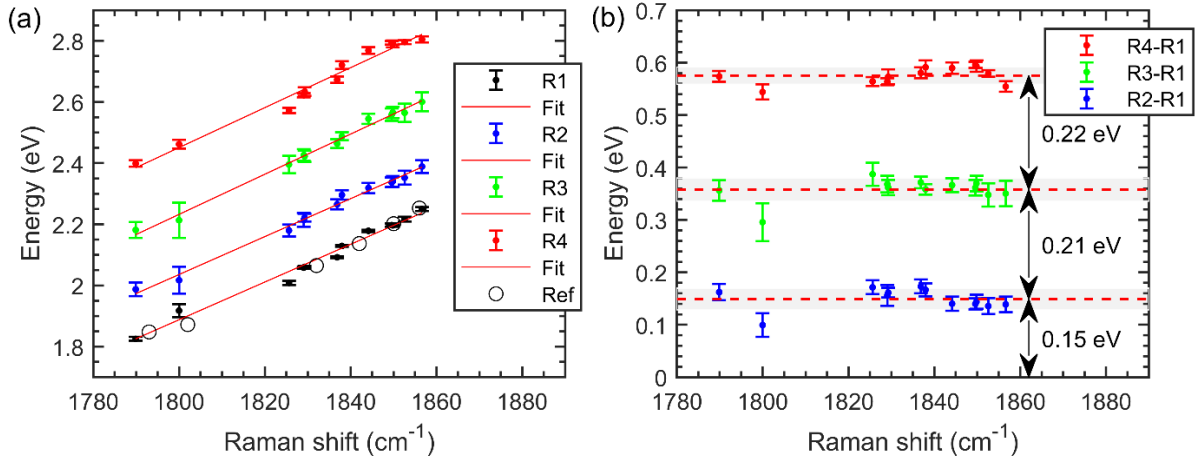

Figure 4: (a) Energy of the four Raman resonances R1-R4 of the RRPs as a function of the Raman frequency of the C-mode and corresponding best linear fits through the data points (red lines). The black circles show the results obtained for the optical gaps, as reported in Ref. [6]. (b) Energy difference between the second, third, and fourth Raman resonances (R2-R4) and the first Raman resonance (R1). The red horizontal lines correspond to the average value and the grey shaded areas, to the standard deviation. The double-sided arrows represent the average energy spacing between consecutive Raman resonances.

In Figure 4a, the energies of the fitted Raman resonances (denoted as R1-R4, with R1 corresponding to the optical gap) are plotted as a function of the corresponding Raman frequency for each fitted Raman frequency component of the C-mode. The energy of each of the components appears to follow a linear relation with the Raman frequency of the C-mode $\nu_C$:

$$E_R = A\nu_C + B$$

with $A=(6.17\pm0.62)\times10^{-3}$ eV/cm$^{-1}$ and $B=(-9.22\pm1.13)$ eV for R1, which is in perfect agreement with the results reported in Ref. [6]. For the second, third, and fourth Raman resonances, A= (6.20±0.42), (6.57±0.40), (6.53±0.75)×10$^{-3}$ eV/cm$^{-1}$, respectively.

Figure 4b shows the differences in energy between the second (R2), third (R3), and fourth (R4) Raman resonances and the optical gap (R1); the values remain constant as a function of Raman shift. The average values of the difference in energy between consecutive Raman resonances is equal to (0.15±0.02), (0.21±0.03), and (0.22±0.03) eV. Interestingly, the spacing in energy between the second and the first Raman resonances (0.15 eV) is significantly lower than the phonon energy in the ground electronic state, which corresponds to 0.22-0.23 eV from the corresponding Raman frequencies (1780-1850 cm$^{-1}$). Thus, interestingly, the resonances are not separated by the same energy differences, a feature that would be expected for a typical vibronic series of peaks.

It is worth noting that, in the absorption spectra of short LCCs, a variation is also observed in the energy differences between the absorption peaks.[14,22] In fact, it is only in the approximation of an ideal harmonic oscillator that the energy spacing between consecutive vibrational states is constant; in a real potential, such as the Morse potential, the energy difference between vibrational levels decreases with increasing energy due to anharmonic effects. Here, surprisingly, we observe the opposite trend as the energy difference between the first and second resonances (R2-R1) is significantly smaller than those of the subsequent resonances. This result suggests the presence of multiple electronic resonances in the RRPs. On the other hand, the differences in energy between the third and second (R3-R2) and the fourth and third (R4-R3) resonances, match well the observed phonon energies; these resonances could thus originate from a vibronic series. We could therefore hypothesize that the four resonances we observe correspond to two electronic transitions, with the first electronic transition corresponding to the optical gap, the second resonance corresponding to the second optical transition, and the third and fourth resonances originating from a vibronic series of the second optical transition. A recent work by Zirzlmeier *et al.* suggested that the optical gap of LCCs does not involve the lowest excited states, which are dark excitonic states.[22] These dipole-forbidden states become weakly allowed in relatively short LCCs and can then be detected by absorption spectroscopy.

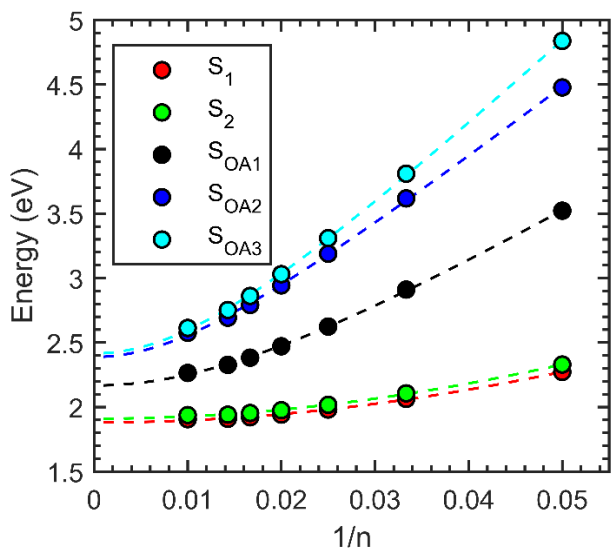

Figure 5: Computed excited-state energies as a function of inverse chain length (1/n) at a dielectric constant of 4. $S_1$ and $S_2$ are the first and second singlet excited states, respectively; $S_{OA1}$, $S_{OA2}$, and $S_{OA3}$ are the first, second, and third optically allowed transitions, respectively. The excited-state energies in an infinite chain are extrapolated by fitting the data with a hyperbolic function described in the SI.5.

In order to verify our hypothesis regarding the nature of the Raman resonances in encapsulated LCCs, we performed TD-DFT calculations for LCCs up to 100 C-atoms; the influence of the CNT host on the electronic structure of the LCCs was modeled via the implicit consideration of a dielectric environment. The results are shown in Figure 5 for a dielectric constant $\varepsilon=4$. Our calculations show that the oscillator strengths (f) of the $S_0 \rightarrow S_1$ and $S_0 \rightarrow S_2$ transitions are vanishingly small (see Table S1 in the SI), which means that these two states are dark states, a result in line with previous studies[22]. As seen from Figure 5, the energies of these two states converge in the case of very long chains, such that it is possible to extrapolate the values to the infinite limit, which can then be compared with the ultralong LCCs inside the CNTs. Depending upon the extrapolation approach we choose (*i.e.*, a linear function or a hyperbola function), the first optically allowed state in an infinite chain, $S_{OA1}$, is calculated to appear some 0.10-0.30 eV above the $S_1$ state. The energies of the next two optically allowed electronic states, $S_{OA2}$ and $S_{OA3}$, converge to the same value in infinite chains, *ca*. 0.14-0.22 eV above $S_{OA1}$ (we note that the excited-state energies are somewhat sensitive to the choice of the dielectric constant value, see Tables S1 and S2 in the SI). These findings thus suggest that the R2 resonance observed in the Raman spectra has an electronic origin. In the case of the $C_{100}H_2$ chain, we calculate that the ratio of the combined oscillator strength of the $S_0 \rightarrow S_{OA2}$ and $S_0 \rightarrow S_{OA3}$ transitions over that of the $S_0 \rightarrow S_{OA1}$ transition is about 0.1, a value in agreement with the experimentally observed ratio of the R2/R1 intensities (0.18±0.08).

In order to evaluate whether the high-energy Raman resonances could be due to vibrational overtones, we computed the electron-vibration couplings (Huang-Rhys factors) for the $S_0 \rightarrow S_{OA1}$, $S_0 \rightarrow S_{OA2}$, and $S_0 \rightarrow S_{OA3}$ transitions (Table S3 in the SI). Due to the complexity of these calculations, the longest chain we were able to consider is the $C_{30}H_2$ chain. The results show that the vibrational mode that exhibits the most intense Raman intensity possesses the largest Huang-Rhys factor (S); the estimated S values for this mode are 0.38, 0.55, and 0.27 for the $S_0 \rightarrow S_{OA1}$, $S_0 \rightarrow S_{OA2}$, and $S_0 \rightarrow S_{OA3}$ transitions, respectively; moreover, in the case of the $S_0 \rightarrow S_{OA3}$ transition, there appears an additional Raman mode with S=0.19. From the relative intensities of the last three Raman resonances (R2, R3 and R4), the Huang-Rhys factors of the transition $S_0 \rightarrow S_{OA2}$ can be experimentally estimated to be on average equal to (1.82±0.14), (see Figure S1 in the SI). The larger S-values for the second optical transition and the two-component third optical transition agree with the hypothesis drawn from our experimental data that the vibronic side-bands manifest more strongly in the higher-energy Raman resonances (however, it should be borne in mind that the ratio between overtone and fundamental intensities is not defined only by S and can depend on several factors, such as the interference between different transitions, temperature and dephasing effects).[40]

## 4. Conclusion

To conclude, in this work we studied the excited states beyond the optical gap of ultralong LCCs encapsulated inside DWCNTs, representing confined carbyne. The four Raman resonances observed in the RRPs of the C-mode of the LCCs are interpreted to come from the optical gap and a second electronic excited state, followed by an intense vibronic series. This interpretation is confirmed by DFT calculations that show both the presence of two optically allowed electronic states separated by 0.15-0.22 eV in the limit of an infinite chain and a strong electron-phonon coupling in the second electronic state, which results in the corresponding intense overtones.

**Acknowledgements**

This research received funding from the European Research Council through an ERC Starting Grant No. 679841 (ORDERin1D) which was granted to S.C. and partially funded the PhD research of M.M., the University of Antwerp Research Fund (BOF-DOCPRO4) that provided M.M. with a PhD grant and the Fund for Scientific Research Flanders (FWO) through projects No. G040011N, G021112N, and G036618N. The work at the University of Arizona was supported by the College of Science. L.S. acknowledges the financial support from the National Natural Science Foundation of China (51902353) and the Guangdong Basic and Applied Basic Research Foundation (2019A1515011227). We thank Takeshi Saito for supplying the eDIPS SWCNTs.

**CRediT author statement**

**Miles Martinati**: Investigation, Formal Analysis, Writing – Original Draft, Methodology. **Wim Wenseleers**: Conceptualization, Methodology, Software, Validation, Investigation, Resources, Writing – Review & Editing, Supervision, Funding acquisition. **Lei Shi**: Conceptualization, Methodology, Validation, Investigation, Resources, Writing – Review & Editing, Funding acquisition. **Saied Md Pratik**: Methodology, Formal Analysis, Software, Writing – Original Draft. **Philip Rohringer**: Investigation, Methodology, Writing-Review & Editing. **Weili Cui:** Resources, Writing-Review & Editing. **Thomas Pichler**: Conceptualization, Resources, Writing-Review & Editing, Funding acquisition. **Veaceslav Coropceanu**: Methodology, Formal Analysis, Writing – Review & Editing. **Jean-Luc Brédas**: Supervision, Writing – Review & Editing, Funding acquisition. **Sofie Cambré**: Conceptualization, Methodology, Software, Validation, Investigation, Resources, Writing – Review & Editing, Supervision, Funding acquisition.

**Appendix A. Supplementary data**

Supplementary data associated with this article can be found in the online version, at ... The supplementary data comprises: (1) TD-DFT-calculated energies and oscillator strengths of the excited states of the LCCs and TD-DFT-calculated Huang-Rhys factors **S** calculated for the lowest three optically allowed singlet excited states of a LCCs of 30 C-atoms, (2) Raman intensity calibration with the Silicon sample, (3) experimentally derived Huang-Rhys factor of the LCCs from the RRPs, (4) example of a fit of the RRP considering the last three Raman resonances as a vibronic series of the same electronic state, (5) fit function for the extrapolation of the excited state energies.

SUPPORTING INFORMATION

# Electronic structure of confined carbyne from joint wavelength-dependent resonant Raman spectroscopy and density functional theory investigations

*Miles Martinati[a], Wim Wenseleers[a*], Lei Shi[b], Saied Md Pratik[c], Philip Rohringer[a,d], Weili Cui[d], Thomas Pichler[d], Veaceslav Coropceanu[c], Jean-Luc Brédas[c], Sofie Cambré[a*].*

[a] Nanostructured and Organic Optical and Electronic Materials group, Physics Department, University of Antwerp, Universiteitsplein 1, B-2610 Antwerp, Belgium

[b] State Key Laboratory of Optoelectronic Materials and Technologies, Nanotechnology Research Center, Guangzhou Key Laboratory of Flexible Electronic Materials and Wearable Devices, School of Materials Science and Engineering, Sun Yat-sen University, Guangzhou 510275, P. R. China

[c] Department of Chemistry and Biochemistry, The University of Arizona, Tucson, Arizona 85721-0088, United States

[d] Faculty of Physics, University of Vienna, Strudlhofgasse 4, A-1090 Vienna, Austria

* Correspondence should be addressed to wim.wenseleers@uantwerp.be (Wim Wenseleers) and sofie.cambre@uantwerpen.be.

## SI.1 Additional data from the TD-DFT calculations

**Table S1**: TD-DFT-calculated energies (E, in eV) and oscillator strengths (f, in parentheses) of the lowest three (dark) excited states and the first three optically allowed excited states, as a function of n in $C_nH_2$ chains, when considering an implicit dielectric constant of 4.

| n | States→ | $S_1$ | $S_2$ | $S_3$ | $S_{OA1}$ | $S_{OA2}$ | $S_{OA3}$ |
|---|---|---|---|---|---|---|---|
| 20 | | 2.27 (0.00) | 2.33 (0.00) | 2.33 (0.00) | 3.52 (7.65) | 4.48 (0.55) | 4.84 (1.64) |
| 30 | | 2.06 (0.00) | 2.11 (0.00) | 2.11 (0.00) | 2.91 (10.1) | 3.62 (1.00) | 3.81 (2.61) |
| 40 | | 1.98 (0.00) | 2.02 (0.00) | 2.02 (0.00) | 2.62 (12.6) | 3.19 (1.56) | 3.31 (2.96) |
| 50 | | 1.94 (0.00) | 1.98 (0.00) | 1.98 (0.00) | 2.47 (15.3) | 2.94 (2.18) | 3.03 (2.74) |
| 60 | | 1.92 (0.00) | 1.95 (0.00) | 1.95 (0.00) | 2.38 (18.4) | 2.79 (2.64) | 2.86 (2.31) |
| 70 | | 1.91 (0.00) | 1.94 (0.00) | 1.94 (0.00) | 2.33 (21.6) | 2.69 (2.79) | 2.75 (1.96) |
| 100 | | 1.91 (0.00) | 1.94 (0.00) | 1.94 (0.00) | 2.26 (32.6) | 2.58 (2.01) | 2.61 (1.82) |

**Table S2**: TD-DFT-calculated energies (E, in eV) of the lowest two (dark) excited states and the first three optically allowed excited states, as a function of n in $C_nH_2$ chains, when considering an implicit dielectric constant of 2.

| n   | $S_1$ | $S_2$ | $S_{OA1}$ | $S_{OA2}$ | $S_{OA3}$ |
|-----|-------|-------|-----------|-----------|-----------|
| 20  | 2.36  | 2.43  | 3.71      | 4.88      | 5.32      |
| 30  | 2.16  | 2.21  | 3.11      | 4.01      | 4.27      |
| 40  | 2.08  | 2.13  | 2.84      | 3.60      | 3.78      |
| 50  | 2.04  | 2.09  | 2.69      | 3.38      | 3.51      |
| 60  | 2.03  | 2.07  | 2.61      | 3.26      | 3.37      |
| 70  | 2.02  | 2.06  | 2.56      | 3.18      | 3.28      |
| 100 | 2.07  | 2.10  | 2.59      | 3.12      | 3.29      |

**Table S3**: Huang-Rhys factors **S** calculated in the lowest three optically allowed singlet excited states of $C_{30}H_2$: $S_{OA1}$, $S_{OA2}$, and $S_{OA3}$.

|    | Frequency (cm$^{-1}$) | S in state $S_{OA1}$ | S in state $S_{OA2}$ | S in state $S_{OA3}$ |
|----|-----------------------|----------------------|----------------------|----------------------|
| Ag | 136                   | 0.06                 | 0.06                 | 0.00                 |
| Ag | 405                   | 0.00                 | 0.00                 | 0.02                 |
| Ag | 667                   | 0.00                 | 0.00                 | 0.00                 |
| Ag | 919                   | 0.00                 | 0.00                 | 0.00                 |
| Ag | 1155                  | 0.00                 | 0.00                 | 0.00                 |
| Ag | 1367                  | 0.00                 | 0.00                 | 0.00                 |
| Ag | 1537                  | 0.00                 | 0.00                 | 0.00                 |
| Ag | 2046                  | 0.38                 | 0.55                 | 0.27                 |
| Ag | 2101                  | 0.00                 | 0.00                 | 0.00                 |
| Ag | 2145                  | 0.00                 | 0.00                 | 0.00                 |
| Ag | 2209                  | 0.01                 | 0.00                 | 0.19                 |
| Ag | 2218                  | 0.00                 | 0.00                 | 0.05                 |
| Ag | 2282                  | 0.00                 | 0.00                 | 0.00                 |
| Ag | 2297                  | 0.00                 | 0.05                 | 0.01                 |
| Ag | 2316                  | 0.00                 | 0.00                 | 0.00                 |
| Ag | 3487                  | 0.00                 | 0.00                 | 0.00                 |

## SI.2 Raman intensity calibration with the Silicon sample

The dependence of the Raman intensity $I$ on the laser excitation energy $E_L$ can be expressed as

$$I(E_L) \propto I_0 \frac{E_S^4}{E_L} |\alpha_{ij}(E_L)|^2$$

with $I_0$ the intensity of the incoming light focused on the sample, $E_S$ the energy of the scattered light and $\alpha_{ij}(E_L)$ the Raman polarizability calculated between the two generic states $ij$. As discussed in the main text, the Raman intensity of the LCC samples at different laser excitation energies are calibrated by dividing the Raman spectra by the intensity of the Raman mode at 520 cm$^{-1}$ of Silicon, measured before and after every measurement:

$$\frac{I^{LCC}(E_L)}{I^{Si}(E_L)} = \left(\frac{E_S^{LCC}}{E_S^{Si}}\right)^4 \frac{|\alpha_{ij}^{LCC}(E_L)|^2}{|\alpha_{ij}^{Si}(E_L)|^2}$$

In this way we can correct over the different intensities $I_0$ of the incoming light on the sample at different laser excitation energies. However, to obtain $|\alpha_{ij}^{LCC}(E_L)|^2$, the experimental data needs to be corrected by the factor $\left(\frac{E_S^{LCC}}{E_S^{Si}}\right)^4$ because at the same $E_L$ the scattered light of the Silicon Raman mode (520 cm$^{-1}$) is different compared to the scattered light of the LCC C-mode (1780-1850 cm$^{-1}$). Furthermore, the experimental data are corrected by the $E_L$ dependence of the Silicon polarizability $|\alpha_{ij}^{Si}(E_L)|^2$, which monotonously increases with laser energy (without sharp resonances) in our laser energy range [2].

The Silicon polarizability $|\alpha_{ij}^{Si}(E_L)|^2$ used in this work to correct the experimental data was estimated by Renucci *et al.* in Ref. [2], combining experimental data (calibrated in their turn using CaF$_2$) and theory. It is important to note that in the work of Renucci *et al.,* the experimental data do not cover our entire laser excitation energy range and thus the Raman cross section is determined with the following polynomial extrapolation:

$$|\alpha_{ij}^{Si}(E_L)|^2 = 8.62 + 1.88 E_L^2 + 0.25 E_L^4 + 0.03 E_L^6 \text{ (eV)}.$$

Note that this extrapolation is expected to become less accurate above 2.7 eV, where the theoretical curve deviates significantly from the experimental data.[2]

## SI.3 Estimation of the Huang-Rhys factors

Figure S1 shows the integrated areas A2-A4 of the Raman resonances R2-R4 normalized by the integrated area A1 of the first Raman resonance R1, as a function of the Raman frequencies of the C-mode. The mean value of the intensity ratio of the second versus the first resonance, *i.e.* A2/A1=(0.18±0.08), agrees with the calculated ratio between the oscillator strengths of the second and first Raman resonances (see Table S1). Comparing the relative amplitudes of the Raman resonances R2-R4, we can estimate the Huang-Rhys factors $S$ of the vibronic series as follows[1]:

$$(A3/A1)/(A2/A1) = |S - 1|^2 \qquad (1)$$

$$(A4/A1)/(A3/A1) = \left|\frac{S}{2}\frac{(S-2)}{(S-1)}\right|^2 \qquad (2)$$

As shown in Figure S1, the relative integrated areas vary a lot for different components of the C-mode and the corresponding mean values are affected by large error bars. This is mainly due to the difficulty to determine the exact amplitude of the resonances because R1 and R2 are largely overlapping and on the other hand R4 has in general a very low intensity. However, from these data we can estimate the HR factors to be on average equal to $S$=(1.82±0.14). This value is significantly higher compared to the calculated value (0.55, see main text) for the optical allowed transition $S_{OA2}$, but qualitatively confirms that this transition has the higher HR factor.

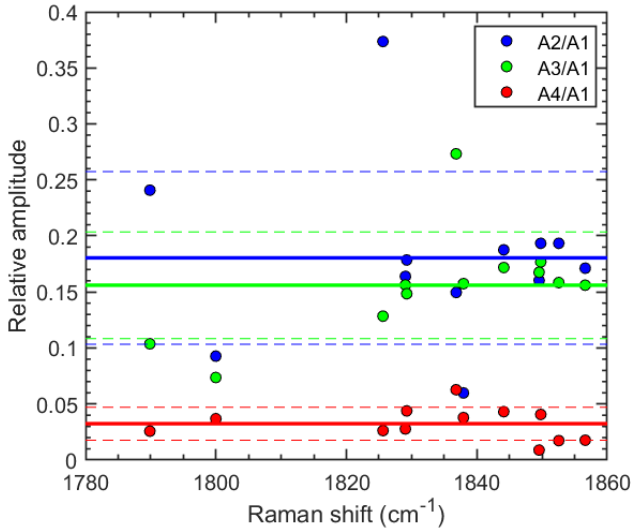

**Figure S1**: Integrated areas A2-A4 of the Raman resonances R2-R4 normalized by the integrated area A1 of the first Raman resonance R1, and plotted as a function of the Raman frequencies of the C-mode obtained from the fits of the RRPs. The horizontal lines correspond to the mean values while the dashed lines represent the standard deviations for each of the three Raman resonances. The large variations of these ratios are mainly originating from the difficulty to determine the exact amplitude of the second Raman resonance R2, which is largely overlapping with the much more intense first Raman resonance R1.

## SI.4 Example fit of the vibronic series in the RRP of the C-modes

According to the DFT calculations and our experimental results, the RRP of the C-modes should be fitted taking into account two different electronic states, *i.e.* $S_{OA1}$ and $S_{OA2}$, and a vibronic series associated to the second electronic state $S_{OA2}$. As such the fit function becomes:

$$I(E_L) = \left|\frac{M_1}{E_L - E_1 - i\Gamma_1}\right|^2 + \left|\sum_{j=0,1,2} \frac{M_{2j}}{E_L - (E_2 + jE_{ph}) - i\Gamma_{2j}}\right|^2 \qquad (3)$$

where the first term corresponds to the first Raman resonance (R1, $S_{OA1}$) and the second term represents the vibronic series of $S_{OA2}$ (resonances R2-R4). Here, the fit parameters are the energies ($E_1$, $E_2$) of the two electronic transitions ($S_{OA1}$, $S_{OA2}$), $M_1$ and $M_{2j}$ are the amplitudes of the four Raman resonances, $\Gamma_1$ and $\Gamma_{2j}$ are the corresponding broadening terms and $E_{ph}$ is the energy of the phonon in the excited state, which can in theory be different from the Raman frequency of the C-mode (phonon energy in the ground state). As can be observed in Figure S2, which shows an example fit of the 1844.1 cm$^{-1}$ C-mode in Sample 2, this fit function can also nicely fit the RRPs, further supporting the origin of the different RRP resonances.

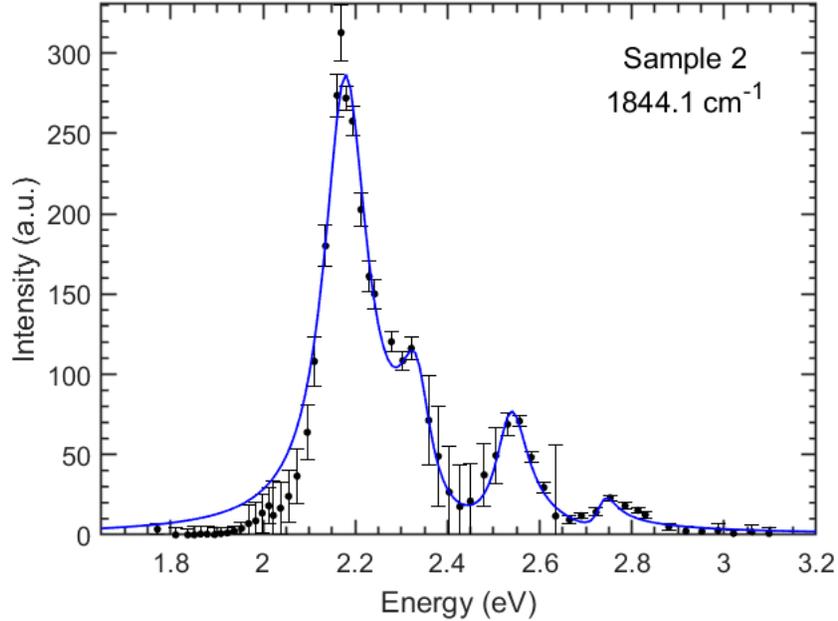

**Figure S2**: Fit of the RRP of the C-mode component at 1844.1 cm$^{-1}$ of Sample 2, with Eq. (3).

## SI.5 Fit function for the extrapolation of the excited state energies

The TD-DFT calculated values of the excited state energies as a function of the inverse number of C-atoms ($n$) are fitted with the following function:

$$f(x) = a + \sqrt{b(2c + bx^2)}$$

where $x = 1/n$, while $a$, $b$ and $c$ are the fitting parameters.

## References:

[1] A.M. Kelley, The Journal of Physical Chemistry A. 117 (2013) 6143–6149. https://doi.org/10.1021/jp400240y

[2] J.B. Renucci, R.N. Tyte, M. Cardona, Resonant Raman scattering in silicon, Physical Review B. 11 (1975). https://doi.org/10.1103/PhysRevB.11.3885.